# A Comparative Study on the Performance of the Top DBMS Systems

Youssef Bassil

LACSC – Lebanese Association for Computational Sciences
Registered under No. 957, 2011, Beirut, Lebanon

**Abstract**

*Database management systems are today's most reliable mean to organize data into collections that can be searched and updated. However, many DBMS systems are available on the market each having their pros and cons in terms of reliability, usability, security, and performance. This paper presents a comparative study on the performance of the top DBMS systems. They are mainly MS SQL Server 2008, Oracle 11g, IBM DB2, MySQL 5.5, and MS Access 2010. The testing is aimed at executing different SQL queries with different level of complexities over the different five DBMSs under test. This would pave the way to build a head-to-head comparative evaluation that shows the average execution time, memory usage, and CPU utilization of each DBMS after completion of the test.*

## 1. Introduction

DBMS short for database management system plays a major role in most real-world projects that require storing, retrieving, and querying digital data. For instance, dynamic websites, accounting information systems, payroll systems, stock management systems all rely on internal databases as a container to store and manage their data [1]. Many software development firms are today developing and producing DBMS systems that cost between zero dollars in case of free and open-source DBMSs, and thousands of dollars in case of proprietary DBMSs. In particular, each DBMS is characterized by a set of diverse functional and non-functional features and specs each having their advantages and disadvantages. One of which is performance which determines how fast a DBMS can process and execute queries. This paper presents a comparative study from a performance perspective between five different DBMSs available today on the market. They are namely MS SQL Server 2008 [2], Oracle 11g [3], IBM DB2 [4], MySQL 5.5 [5], and MS Access 2010 [6]. For this reason, several SQL queries with different level of complexities were crafted and tested against all these well-known DBMSs. Additionally, a performance benchmark was used to measure the execution time of every executed SQL query, in addition to CPU utilization, memory usage, virtual memory usage, and threads count. In due course, a head-to-head comparison was drawn, which exhibits the differences in performance between the different DBMSs under test.

## 2. Background

This section discusses the history, versions, and features of the different DBMSs under test. They are respectively MS SQL Server 2008, Oracle 11g, IBM DB2, MySQL 5.5, and MS Access 2010.

### 2.1. MS SQL Server 2008

Microsoft SQL Server is a relational database management system (RDBMS) produced by Microsoft. Its primary query language is Transact-SQL, an implementation of the ANSI/ISO standard Structured Query Language (SQL) used by both Microsoft and Sybase. Microsoft SQL Server supports atomic, consistent, isolated, and durable transactions. It includes support for database mirroring and clustering. An SQL server cluster is a collection of identically configured servers, which help distribute the workload among multiple servers. SQL server also supports data partitioning for distributed databases, in addition to database mirroring which allows the creation of mirrors of database contents, along with transaction logs, on another instance of SQL Server, based on certain predefined triggers [7].

## 2.2. Oracle 11g

Oracle Database (commonly referred to as Oracle RDBMS or simply as Oracle), is a relational database management system (RDBMS) released by Oracle Corporation, and it comprises at least one instance of the application, along with data storage. An instance comprises a set of operating system processes and memory structures that interact with the storage. In addition to storage, the database consists of online redo logs which hold the transactional history. Processes can in turn archive the online redo logs into archive logs, which provide the basis for data recovery and for some forms of data replication. The Oracle RDBMS stores data logically in the form of table-spaces and physically in the form of data files. At the physical level, data files comprise one or more data blocks, where the block size can vary between data files. Oracle features data dictionary, indexes, and clusters. Versions Subsequent to 10g, introduced grid computing capabilities in which an instance application can use CPU resources from another node in the grid [8].

## 2.3. IBM DB2

DB2 is one of IBM's lines of relational database management system which runs on Unix, Windows, or Linux server machines. DB2 can be administered from either a command-line or a GUI interface. The command-line interface requires more knowledge of the product but can be more easily scripted and automated. The GUI is a multi-platform Java client that contains a variety of wizards suitable for novice users. DB2 supports both SQL and XQuery. DB2 has native implementation of XML data storage, where XML data is stored as XML for faster access using XQuery. DB2 also supports integration into the Eclipse and Visual Studio .NET integrated development environments. An important feature of DB2 DBMS is the error processing in which SQL communications area structure is used within the DB2 program to return error information to the application program after every API call for an SQL statement [9].

## 2.4. MySQL 5.5

MySQL  is a free, open-source, multithreaded, and multi-user SQL database management system which has more than 10 million installations. The basic program runs as a server providing multi-user access to a number of databases. MySQL includes a broad subset of ANSI SQL 99, as well as extensions, cross-platform support, stored procedures, triggers, cursors, updatable views, and X/Open XA distributed transaction processing support. Moreover, it supports two phase commit engine, independent storage engines, SSL support, query caching, replication with one master per slave, many slaves per master, embedded database library, and ACID compliance using the InnoDB cluster engines [10].

## 2.5. MS Access 2010

Microsoft Office Access, previously known as Microsoft Access, is a relational database management system from Microsoft which combines the relational Microsoft Jet Database Engine with a graphical user interface and software development tools. It is a member of the 2010 Microsoft Office system. One of the benefits of Access from a programmer's perspective is its relative compatibility with SQL queries. Unlike a complete RDBMS, the Jet Engine lacks database triggers and stored procedures. Notwithstanding, it provides a special syntax that allows creating queries with parameters, in a way that looks like creating stored procedures, but these procedures are limited to one statement per procedure. Microsoft Access does allow forms to contain code that is triggered as changes are made to the underlying table, and it is common to use pass-through queries and other techniques in Access to run stored procedures in RDBMSs that support these. MS Access is used by small businesses, within departments of large corporations, and by hobby programmers to create ad hoc customized desktop systems for handling the creation and manipulation of data. Some professional application developers use Access for rapid application development, especially for the creation of prototypes and standalone applications that serve as tools for on-the-road salesmen [11].

## 3. Testing and Evaluation

### 3.1. DBMSs under Test

There are typically five DBMSs under test, four of which are client/server DBMSs, suitable for building medium and large scale databases, and one standalone DBMS suitable for creating small scale ad-hoc databases. They are respectively MS SQL Server 2008, Oracle 11g, IBM DB2, MySQL 5.5, and MS Access 2010. MS Access is the only non-client/server DBMS.

## 3.2. Testing Platform

The testing is carried out on a Dual-Processor, Intel Xeon E5649, 6x2 Cores, processor, clocked at 2.53GHz with 32GB of random access memory (RAM) and 2TB of secondary storage capacity. The operating system is MS Windows Server 2008, 64-bit.

## 3.3. Tester

The tester is a computer application developed using C#.NET under the .NET Framework 4.0. It performs two tasks: The first is to automatically populate the database tables with 1,000,000 rows prior to test execution. The second is to execute the actual SQL queries. Figure 1 shows the main GUI interface of the tester.

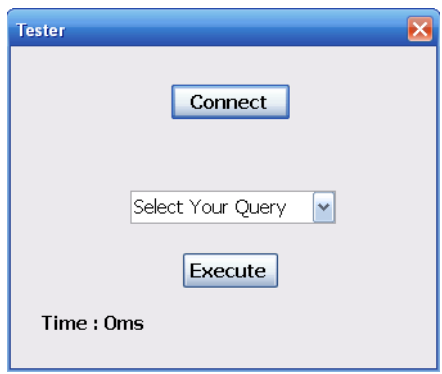

Figure 1 – Tester Interface

## 3.4. Benchmarking

The tester implements a built-in timer to measure the execution time in milliseconds, from the start of the execution of a particular SQL query until it finishes up. Concerning memory consumption and utilization, the MS Windows Task Manager (WTM) tool is used which is already shipped with all versions of MS Windows operating systems [12]. Figure 2 shows the interface of the WTM tool

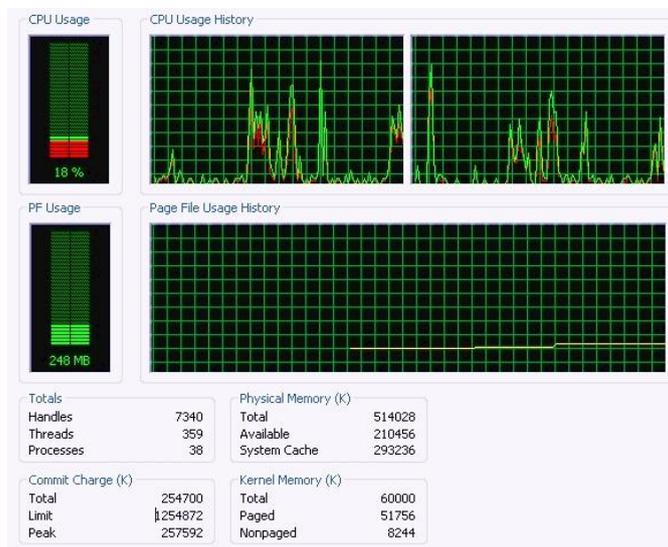

Figure 2 – WTM Main Interface

## 3.5. Database Design

Essentially, the database to be tested over all the different DBMSs comprises fifteen distinct relations or tables associated together by means of relationships. It is a relational model database implemented under the different five DBMSs under test. The database fits a business retail system. It includes a front end system for creating invoices, receipts, and purchase orders and a back end system to manage the items stock. Figure 3 depicts the logical design of the database under test.

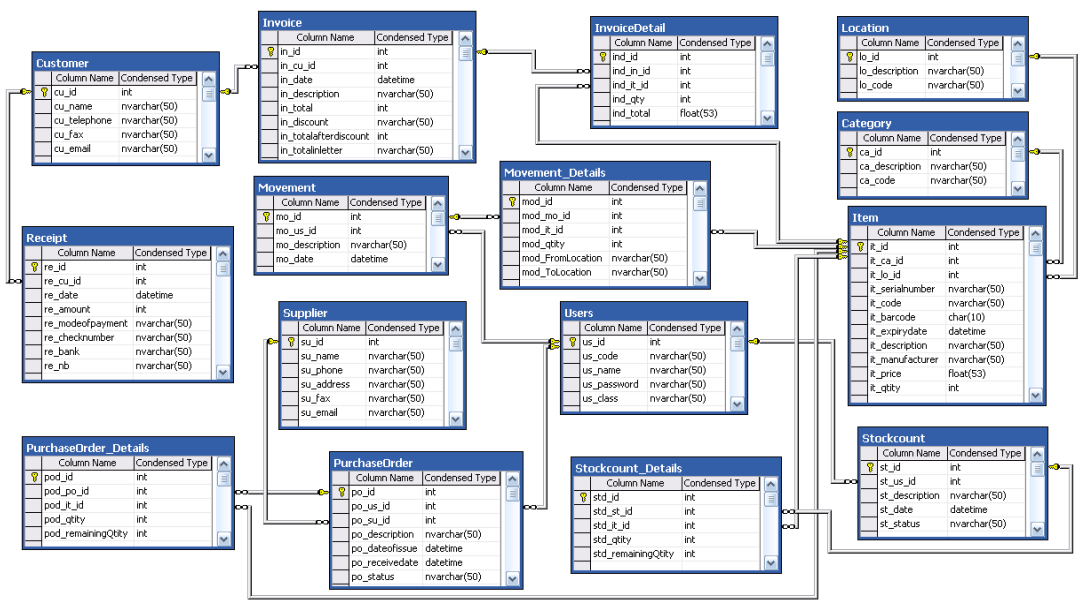

Figure 3 - Logical Design of the Database

## 4. The Testing Process

Different SQL queries were executed over the different five DBMSs under test. In fact, these queries have different level of complexity; they range from simple type to very complex type. It is worth noting that all five databases are populated with dummy 1,000,000 records of data prior to starting the testing process.

**Query #1**

This is a *very simple* query whose task is to retrieve rows without any conditions or joins:

SELECT * FROM Item;

|  | Execution Time | CPU Utilization | Memory Utilization | Virtual Memory Utilization | Threads Used |
|---|---|---|---|---|---|
| SQL Server | 18 ms | +3% | + 3MB | + 1MB | + 2 |
| Oracle | 23 ms | +4% | + 7MB | + 1MB | + 6 |
| IBM DB2 | 18 ms | +3% | + 11MB | + 1MB | + 2 |
| MySQL | 19 ms | +3% | + 3MB | + 1MB | + 2 |
| Ms Access | 21 ms | +2% | + 2MB | + 1MB | + 1 |

**Query #2**

This query employs the use of *sophisticated conditions* conjoined with logical operators:

SELECT * FROM Invoice

WHERE Invoice.in_id > 50 AND Invoice.in_date > 1/1/2006 AND Invoice.in_date < 1/1/2007 AND Invoice.in_description LIKE '%ohp%' AND Invoice.in_totalinletter LIKE '%USD' AND Invoice.in_total = Invoice.in_totalafterdiscount AND Invoice.in_total <> 100 OR NOT Invoice.in_cu_id >= 5 AND Invoice.in_id BETWEEN 1 AND 10000 OR Invoice.in_id > 49+1 AND Invoice.in_total+33 <> 5 AND Invoice.in_total = - Invoice.in_totalafterdiscount * 2 ;

|  | Execution Time | CPU Utilization | Memory Utilization | Virtual Memory Utilization | Threads Used |
|---|---|---|---|---|---|
| SQL Server | 124 ms | +9% | + 3MB | + 1MB | + 2 |
| Oracle | 125 ms | +14% | + 7MB | + 1MB | + 6 |
| IBM DB2 | 125 ms | +12% | + 11MB | + 1MB | + 2 |
| MySQL | 126 ms | +12% | + 3MB | + 1MB | + 2 |
| Ms Access | 170 ms | +6% | + 2MB | + 1MB | + 1 |

**JCSCR**

Journal of Computer
Science & Research

## Query #3

This query is used to test *the join* operation between different tables:

SELECT Customer.cu_id , Invoice.in_id , InvoiceDetail.ind_qty , Item.it_serialnumber , Movement.mo_description , Movement_Details.mo_it_id , Users.us_id , Users.us_code , PurchaseOrder.po_description , Supplier.su_name FROM Customer , Invoice , InvoiceDetail , Item , Movement , Movement_Details , Users , PurchaseOrder , Supplier
WHERE Supplier.su_name = "Mike" AND Customer.cu_id = Invoice.in_cu_id AND InvoiceDetail.ind_in_id = Invoice.in_id AND InvoiceDetail.ind_it_id = Item.it_id AND Movement_Details.mod_mo_id = Movement.mo_id AND Movement.mo_us_id = Users.us_id AND PurchaseOrder.po_us_id =Users.us_id AND PurchaseOrder.po_us_id = Users.us_id AND PurchaseOrder.po_su_id AND Supplier.su_name ;

| | Execution Time | CPU Utilization | Memory Utilization | Virtual Memory Utilization | Threads Used |
|---|---|---|---|---|---|
| SQL Server | 123 ms | +3% | + 33MB | + 3MB | + 2 |
| Oracle | 122 ms | +4% | + 37MB | + 4MB | + 6 |
| IBM DB2 | 123 ms | +3% | + 43MB | + 5MB | + 2 |
| MySQL | 126 ms | +3% | + 27MB | + 3MB | + 2 |
| Ms Access | 231 ms | +3% | + 26MB | + 3MB | + 1 |

## Query #4

This query is used to test *the sorting* operation for each DBMS:

SELECT Customer.cu_id , Customer.cu_name , Customer.cu_telephone , Customer.cu_fax , Customer.cu_email FROM Customer ORDER BY Customer.cu_id , Customer.cu_name DESC , Customer.cu_telephone DESC, Customer.cu_fax , Customer.cu_email DESC ;

| | Execution Time | CPU Utilization | Memory Utilization | Virtual Memory Utilization | Threads Used |
|---|---|---|---|---|---|
| SQL Server | 429 ms | +29% | + 3MB | + 1MB | + 2 |
| Oracle | 431 ms | +41% | + 7MB | + 1MB | + 6 |
| IBM DB2 | 423 ms | +38% | + 11MB | + 1MB | + 2 |
| MySQL | 428 ms | +18% | + 3MB | + 1MB | + 2 |
| Ms Access | 440 ms | +17% | + 2MB | + 1MB | + 1 |

## Query #5

The purpose of this query is to test *computational capabilities* of the DBMSs by executing different *arithmetic functions*:

SELECT SUM(Invoice.in_total) , AVG(Invoice.in_totalafterdiscount) , MAX(Invoice.in_total) , COUNT(Customer.cu_id) , SUM(InvoiceDetail.ind_qty) FROM Customer , Invoice , InvoiceDetail WHERE Customer.cu_id = Invoice.in_cu_id AND Invoice.in_id = InvoiceDetail.ind_in_id GROUP BY Invoice.in_id ;

| | Execution Time | CPU Utilization | Memory Utilization | Virtual Memory Utilization | Threads Used |
|---|---|---|---|---|---|
| SQL Server | 777 ms | +54% | + 13MB | + 1MB | + 2 |
| Oracle | 801 ms | +70% | + 16MB | + 2MB | + 6 |
| IBM DB2 | 650 ms | +55% | + 21MB | + 2MB | + 2 |
| MySQL | 732 ms | +35% | + 13MB | + 1MB | + 2 |
| Ms Access | 1234 ms | +33% | + 10MB | + 1MB | + 1 |

## Query #6

This query adds to the previous query conditions after the *HAVING* clause:

SELECT SUM(Invoice.in_total) , AVG(Invoice.in_totalafterdiscount) , MAX(Invoice.in_total) , COUNT(Customer.cu_id) , SUM(InvoiceDetail.ind_qty) FROM Customer , Invoice , InvoiceDetail
WHERE Customer.cu_id = Invoice.in_cu_id AND Invoice.in_id = InvoiceDetail.ind_in_id GROUP BY Invoice.in_id HAVING COUNT(Invoice.in_id)>0 AND SUM(Invoice.in_total) = AVG(Invoice,in_totalafterdiscount) ;

|  | Execution Time | CPU Utilization | Memory Utilization | Virtual Memory Utilization | Threads Used |
|---|---|---|---|---|---|
| SQL Server | 2304 ms | +60% | + 13MB | + 1MB | + 2 |
| Oracle | 2700 ms | +77% | + 16MB | + 2MB | + 6 |
| IBM DB2 | 2001 ms | +61% | + 21MB | + 2MB | + 2 |
| MySQL | 2732 ms | +46% | + 13MB | + 1MB | + 2 |
| Ms Access | 3001 ms | +41% | + 10MB | + 1MB | + 1 |

## Query #7

This query tests the capabilities of each DBMS when inner *nested SELECTs* is used:

SELECT Customer.cu_name FROM Customer WHERE Customer.cu_name = (SELECT Users.us_name FROM Users WHERE Users.us_class = "administrator") AND Customer.cu_fax = (SELECT Supplier.su_fax FROM Supplier WHERE Supplier.su_phone = "123456") AND Customer.cu_email = (SELECT Supplier.su_email FROM Suppliers WHERE Supplier.su_address LIKE "%h%") ;

|  | Execution Time | CPU Utilization | Memory Utilization | Virtual Memory Utilization | Threads Used |
|---|---|---|---|---|---|
| SQL Server | 292 ms | +3% | + 17MB | + 2MB | + 2 |
| Oracle | 290 ms | +4% | + 24MB | + 2MB | + 6 |
| IBM DB2 | 650 ms | +3% | + 27MB | + 3MB | + 2 |
| MySQL | 340 ms | +3% | + 19MB | + 2MB | + 2 |
| Ms Access | 698 ms | +2% | + 15MB | + 1MB | + 1 |

## Query #8

Now comes the ultimate test which will *combine* all previous queries into a single atomic SQL query:

SELECT Customer.cu_id , Invoice.in_id , InvoiceDetail.ind_qty , Item.it_serialnumber , Movement.mo_description , Movement_Details.mo_it_id , Users.us_id , Users.us_code , PurchaseOrder.po_description , Supplier.su_name , SUM(Invoice.in_total) , AVG(Invoice.in_totalafterdiscount) , MAX(Invoice.in_total), COUNT(Customer.cu_id) , SUM(InvoiceDetail.ind_qty) FROM Customer , Invoice , InvoiceDetail , Item , Movement , Movement_Details , Users , PurchaseOrder , Supplier WHERE Invoice.in_id > 50 AND Invoice.in_date > 1/1/2006 AND Invoice.in_date < 1/1/2007 AND Invoice.in_description LIKE '%ohp%' AND Invoice.in_totalinletter LIKE '%USD' AND Invoice.in_total = Invoice.in_totalafterdiscount AND Invoice.in_total <> 100 OR NOT Invoice.in_cu_id >=5 AND Invoice.in_id BETWEEN 1 AND 10000 OR Invoice.in_id > 49+1 AND Customer.cu_name = (SELECT Users.us_name FROM Users WHERE Users.us_class = "administrator") AND Customer.cu_fax = (SELECT Supplier.su_fax FROM Supplier WHERE Supplier.su_phone = "123456") AND Customer.cu_id = Invoice.in_cu_id AND InvoiceDetail.ind_in_id = Invoice.in_id AND InvoiceDetail.ind_it_id = Item.it_id AND Movement_Details.mod_mo_id = Movement.mo_id AND Movement.mo_us_id = Users.us_id AND PurchaseOrder.po_us_id =Users.us_id AND PurchaseOrder.po_us_id = Users.us_id AND PurchaseOrder.po_su_id AND Supplier.su_id ; ORDER BY Customer.cu_id , Customer.cu_name DESC , Invoice.in_id DESC, Users.us_name , Invoice.in_description DESC ; GROUP BY Customer.cu_id , Invoice.in_id , InvoiceDetail.ind_qty , Item.it_serialnumber , Movement.mo_description , Movement_Details.mo_it_id , Users.us_id , Users.us_code , PurchaseOrder.po_description , Supplier.su_name HAVING COUNT(Invoice.in_id)>0 AND SUM(Invoice.in_total) = AVG(Invoice,in_totalafterdiscount) ;

|  | Execution Time | CPU Utilization | Memory Utilization | Virtual Memory Utilization | Threads Used |
|---|---|---|---|---|---|
| SQL Server | 6790 ms | +99% | + 41MB | + 3MB | + 2 |
| Oracle | 8100 ms | +100% | + 51MB | + 4MB | + 6 |
| IBM DB2 | 6071 ms | +99% | + 59MB | + 5MB | + 2 |
| MySQL | 7520 ms | +99% | + 38MB | + 3MB | + 2 |
| Ms Access | 12678 ms | +99% | + 31MB | + 3MB | + 1 |

## Query #9

This query tests the capabilities of the DBMSs under test to execute *UPDATE* statements with complicated conditions:

UPDATE Item SET Item.it_price = Item.it_price * 0.1 AND Item.it_qtity = 10 AND Item.it_description = "TV" WHERE Item.it_id > 10 AND Item.it_expirydate > 1/1/2007 AND Item.it_expirydate < 1/1/2008 AND Item.it_manufacturer = "Philips" OR Item.it_manufacturer = "Sharp" OR Item.it_manufacturer = "Toshiba" ;

|  | Execution Time | CPU Utilization | Memory Utilization | Virtual Memory Utilization | Threads Used |
|---|---|---|---|---|---|
| SQL Server | 45 ms | +7% | + 3MB | + 1MB | + 2 |
| Oracle | 21 ms | +11% | + 7MB | + 1MB | + 6 |
| IBM DB2 | 102 ms | +8% | + 11MB | + 1MB | + 2 |
| MySQL | 52 ms | +8% | + 3MB | + 1MB | + 2 |
| Ms Access | 201 ms | +7% | + 2MB | + 1MB | + 1 |

**Query #10**

This final query executes a *DELETE* query over the selected DBMSs:

DELETE FROM Invoice WHERE Invoice.in_date > 1/1/2006 AND Invoice.in_date < 1/1/2007 AND Invoice.in_description LIKE '%vtt%' AND Invoice.in_totalinletter LIKE '%USD' AND Invoice.in_total = Invoice.in_totalafterdiscount AND Invoice.in_ totalafterdiscount <> 33.1 OR NOT Invoice.in_cu_id >= 5 AND Invoice.in_id BETWEEN 1 AND 10000 OR Invoice.in_id < 71/2 AND Invoice.in_total+33 <> 5 AND Invoice.in_total = Invoice.in_totalafterdiscount − 112 ;

|  | Execution Time | CPU Utilization | Memory Utilization | Virtual Memory Utilization | Threads Used |
|---|---|---|---|---|---|
| SQL Server | 111 ms | +7% | + 3MB | + 1MB | + 2 |
| Oracle | 140 ms | +11% | + 7MB | + 1MB | + 6 |
| IBM DB2 | 160 ms | +8% | + 11MB | + 1MB | + 2 |
| MySQL | 148 ms | +8% | + 3MB | + 1MB | + 2 |
| Ms Access | 182 ms | +7% | + 2MB | + 1MB | + 1 |

## 5. Results & Conclusions

The results of the testing are represented using graphical charts and statistical histograms. Obviously, there is no ultimate winner. The charts clearly show that IBM DB2 is the fastest DBMS, however MS Access has lower CPU utilization than other DBMSs and IBM DB2 is the most DBMS that consumes primary memory. Figure 4 represents the average execution time, Figure 5 represents the average CPU utilization, and Figure 6 represents the average memory utilization.

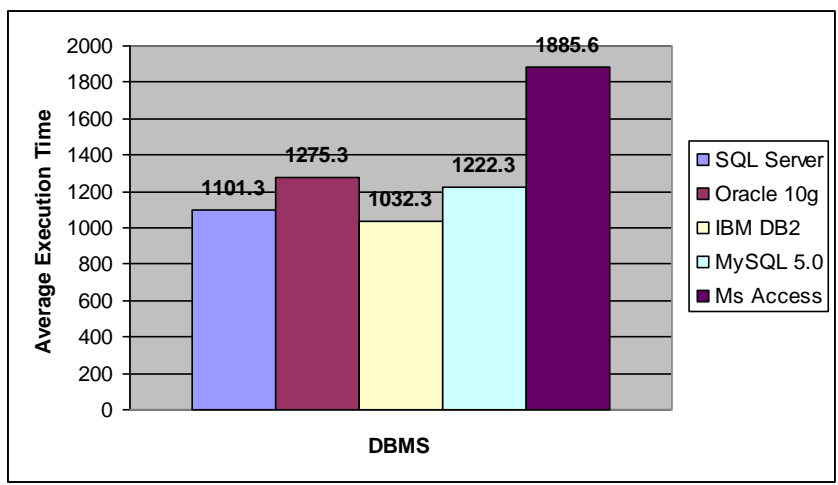

Figure 4 - Average Execution Time

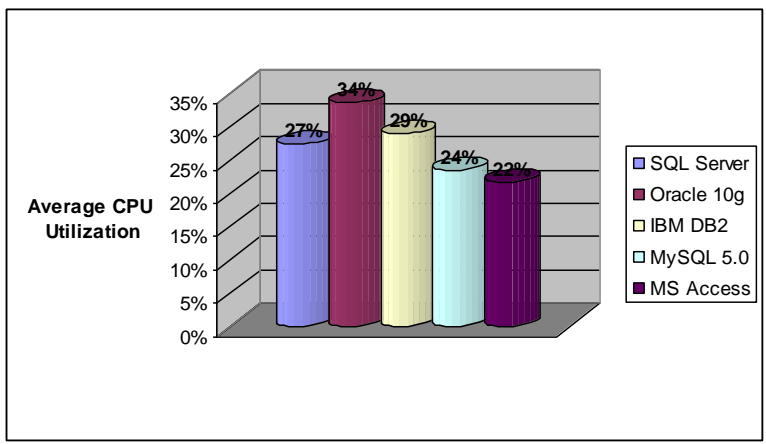

Figure 5 - Average CPU Utilization

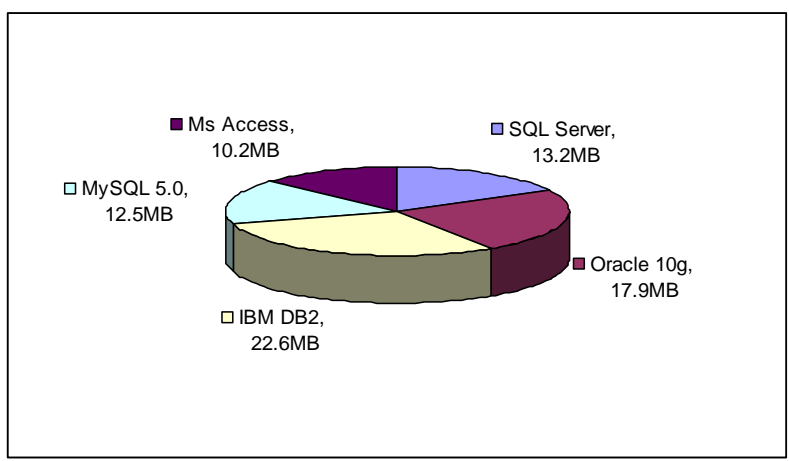

Figure 6 – Average Memory Usage

## Acknowledgment

This research was funded by the Lebanese Association for Computational Sciences (LACSC), Beirut, Lebanon, under the "Evaluation & Performance Research Project – EPRP2012".

## Appendix A

This appendix lists the different DDL queries that were used to build and implement the relational database along with its relationships and constraints.

```
DROP DATABASE IF EXISTS `uniDB`;
CREATE DATABASE ` uniDB ` /*!40100 DEFAULT CHARACTER SET latin1 */;
USE ` uniDB `;
CREATE TABLE `category` (
 `ca_id` int(11) NOT NULL auto_increment,
 `ca_description` varchar(50) default NULL,
 `ca_code` varchar(50) default NULL,
 PRIMARY KEY (`ca_id`)
) ENGINE=InnoDB DEFAULT CHARSET=latin1;

CREATE TABLE `customer` (
 `cu_id` int(11) NOT NULL auto_increment,
 `cu_name` varchar(50) default NULL,
 `cu_telephone` varchar(50) default NULL,
 `cu_fax` varchar(50) default NULL,
 `cu_email` varchar(50) default NULL,
 PRIMARY KEY (`cu_id`)
) ENGINE=InnoDB DEFAULT CHARSET=latin1;

CREATE TABLE `invoice` (
 `in_id` int(11) NOT NULL auto_increment,
 `in_cu_id` int(11) default NULL,
 `in_date` datetime default NULL,
 `in_description` char(50) default NULL,
 `in_total` int(11) default NULL,
 `in_discount` char(50) default NULL,
 `in_totalafterdiscount` int(11) default NULL,
 PRIMARY KEY (`in_id`),
 KEY `in_cu_id` (`in_cu_id`)
) ENGINE=InnoDB DEFAULT CHARSET=latin1 ROW_FORMAT=FIXED;

CREATE TABLE `invoicedetail` (
 `ind_id` int(11) NOT NULL auto_increment,
 `ind_in_id` int(11) default NULL,
 `ind_it_id` int(11) default NULL,
 `ind_qty` int(11) default NULL,
 `ind_total` float(53,10) default NULL,
 PRIMARY KEY (`ind_id`),
 KEY `ind_in_id` (`ind_in_id`),
 KEY `ind_it_id` (`ind_it_id`)
) ENGINE=InnoDB DEFAULT CHARSET=latin1;

CREATE TABLE `item` (
 `it_id` int(11) NOT NULL auto_increment,
 `it_ca_id` int(11) default NULL,
 `it_lo_id` int(11) default NULL,
 `it_serialnumber` char(50) default NULL,
 `it_code` char(50) default NULL,
 `it_barcode` char(10) default NULL,
 `it_expirydate` datetime default NULL,
 `it_description` char(50) default NULL,
 `it_manufacturer` char(50) default NULL,
 `it_price` float(53,10) default NULL,
 `it_qtity` int(11) default NULL,
 PRIMARY KEY (`it_id`),
 KEY `it_ca_id` (`it_ca_id`),
 KEY `it_lo_id` (`it_lo_id`)
) ENGINE=InnoDB DEFAULT CHARSET=latin1 ROW_FORMAT=FIXED;

CREATE TABLE `location` (
```

```
 `lo_id` int(11) NOT NULL auto_increment,
 `lo_description` varchar(50) default NULL,
 `lo_code` varchar(50) default NULL,
 PRIMARY KEY (`lo_id`)
) ENGINE=InnoDB DEFAULT CHARSET=latin1;

CREATE TABLE `movement` (
 `mo_id` int(11) NOT NULL auto_increment,
 `mo_us_id` int(11) default NULL,
 `mo_description` char(50) default NULL,
 `mo_date` datetime default NULL,
 PRIMARY KEY (`mo_id`),
 KEY `mo_us_id` (`mo_us_id`)
) ENGINE=InnoDB DEFAULT CHARSET=latin1 ROW_FORMAT=FIXED;

CREATE TABLE `movement_details` (
 `mod_id` int(11) NOT NULL auto_increment,
 `mod_mo_id` int(11) default NULL,
 `mod_it_id` int(11) default NULL,
 `mod_qtity` int(11) default NULL,
 `mod_fromlocation` char(50) default NULL,
 `mod_tolocation` char(50) default NULL,
 PRIMARY KEY (`mod_id`),
 KEY `mod_it_id` (`mod_it_id`),
 KEY `mod_mo_id` (`mod_mo_id`)
) ENGINE=InnoDB DEFAULT CHARSET=latin1 ROW_FORMAT=FIXED;

CREATE TABLE `purchaseorder` (
 `po_id` int(11) NOT NULL auto_increment,
 `po_us_id` int(11) default NULL,
 `po_su_id` int(11) default NULL,
 `po_description` char(50) default NULL,
 `po_dateofissue` datetime default NULL,
 `po_recievedate` datetime default NULL,
 `po_status` char(50) default NULL,
 PRIMARY KEY (`po_id`),
 KEY `po_us_id` (`po_us_id`),
 KEY `po_su_id` (`po_su_id`)
) ENGINE=InnoDB DEFAULT CHARSET=latin1 ROW_FORMAT=FIXED;

CREATE TABLE `purchaseorder_details` (
 `pod_id` int(11) NOT NULL auto_increment,
 `pod_po_id` int(11) default NULL,
 `pod_it_id` int(11) default NULL,
 `pod_qtity` int(11) default NULL,
 `pod_remainingqtity` int(11) default NULL,
 PRIMARY KEY (`pod_id`),
 KEY `pod_po_id` (`pod_po_id`),
 KEY `pod_it_id` (`pod_it_id`)
) ENGINE=InnoDB DEFAULT CHARSET=latin1;

CREATE TABLE `reciept` (
 `re_id` int(11) NOT NULL auto_increment,
 `re_cu_id` int(11) default NULL,
 `re_date` datetime default NULL,
 `re_amount` int(11) default NULL,
 `re_modeofpayment` varchar(50) default NULL,
 `re_checknumber` varchar(50) default NULL,
 `re_bank` varchar(50) default NULL,
 PRIMARY KEY (`re_id`),
 KEY `re_cu_id` (`re_cu_id`)
) ENGINE=InnoDB DEFAULT CHARSET=latin1;

CREATE TABLE `stockcount` (
```

```
  `st_id` int(11) NOT NULL auto_increment,
  `st_us_id` int(11) default NULL,
  `st_description` varchar(50) default NULL,
  `st_date` datetime default NULL,
  `st_status` varchar(50) default NULL,
  PRIMARY KEY (`st_id`),
  KEY `st_us_id` (`st_us_id`)
) ENGINE=InnoDB DEFAULT CHARSET=latin1;

CREATE TABLE `stockcount_details` (
  `std_id` int(11) NOT NULL auto_increment,
  `std_st_id` int(11) default NULL,
  `std_it_id` int(11) default NULL,
  `std_qtity` int(11) default NULL,
  `std_remainingqtity` int(11) default NULL,
  PRIMARY KEY (`std_id`),
  KEY `std_st_id` (`std_st_id`),
  KEY `std_it_id` (`std_it_id`)
) ENGINE=InnoDB DEFAULT CHARSET=latin1;

CREATE TABLE `supplier` (
  `su_id` int(11) NOT NULL auto_increment,
  `su_name` varchar(50) default NULL,
  `su_phone` varchar(50) default NULL,
  `su_address` varchar(50) default NULL,
  `su_fax` varchar(50) default NULL,
  `su_email` varchar(50) default NULL,
  PRIMARY KEY (`su_id`)
) ENGINE=InnoDB DEFAULT CHARSET=latin1;

CREATE TABLE `users` (
  `us_id` int(11) NOT NULL auto_increment,
  `us_code` varchar(50) default NULL,
  `us_name` varchar(50) default NULL,
  `us_password` varchar(50) default NULL,
  `us_class` varchar(50) default NULL,
  PRIMARY KEY (`us_id`)
) ENGINE=InnoDB DEFAULT CHARSET=latin1;

ALTER TABLE `invoice`
  ADD FOREIGN KEY (`in_cu_id`) REFERENCES `customer` (`cu_id`);

ALTER TABLE `invoicedetail`
  ADD FOREIGN KEY (`ind_it_id`) REFERENCES `item` (`it_id`),
  ADD FOREIGN KEY (`ind_in_id`) REFERENCES `invoice` (`in_id`);

ALTER TABLE `item`
  ADD FOREIGN KEY (`it_ca_id`) REFERENCES `category` (`ca_id`),
  ADD FOREIGN KEY (`it_lo_id`) REFERENCES `location` (`lo_id`);

ALTER TABLE `movement`
  ADD FOREIGN KEY (`mo_us_id`) REFERENCES `users` (`us_id`);

ALTER TABLE `movement_details`
  ADD FOREIGN KEY (`mod_mo_id`) REFERENCES `movement` (`mo_id`),
  ADD FOREIGN KEY (`mod_it_id`) REFERENCES `item` (`it_id`);

ALTER TABLE `purchaseorder`
  ADD FOREIGN KEY (`po_su_id`) REFERENCES `supplier` (`su_id`),
  ADD FOREIGN KEY (`po_us_id`) REFERENCES `users` (`us_id`);

ALTER TABLE `purchaseorder_details`
  ADD FOREIGN KEY (`pod_po_id`) REFERENCES `purchaseorder` (`po_id`),
```

```
  ADD FOREIGN KEY (`pod_it_id`) REFERENCES `item` (`it_id`);

ALTER TABLE `reciept`
  ADD FOREIGN KEY (`re_cu_id`) REFERENCES `customer` (`cu_id`);
ALTER TABLE `stockcount`
  ADD FOREIGN KEY (`st_us_id`) REFERENCES `users` (`us_id`);

ALTER TABLE `stockcount_details`
  ADD FOREIGN KEY (`std_st_id`) REFERENCES `stockcount` (`st_id`),
  ADD FOREIGN KEY (`std_it_id`) REFERENCES `item` (`it_id`);
```